\documentstyle{europhys}

\def\etal{{\hbox{{\it\ et al.\/}\rm :\ }}}

\def\And{{\rm and\ }}

\def\stars{\bigskip\centerline{***}\medskip}

\newif\ifboo \boofalse

\def\Review#1{\boofalse{\it #1},}
\def\Name#1{{\sc #1},}
\def\Vol#1{\ifboo Vol. {\bf #1}\else{\bf #1}\fi}
\def\Year#1{\ifboo #1\else(#1)\fi}
\def\Book#1{\bootrue{\it #1},}
\def\Page#1{\ifboo {\rm p. #1}\else{\rm #1}\fi}

\begin{document}
\euro{}{}{}{}
\Date{}
\shorttitle{J. MAIN \etal SEMICLASSICAL QUANTIZATION BY PAD\'E APPROXIMANT ETC.}
\title{Semiclassical quantization by Pad\'e approximant to\\
 periodic orbit sums}
\author{J\"org Main\inst{1}, Paul A. Dando\inst{2},
    D\v zevad Belki\'c\inst{2} \And Howard S. Taylor\inst{2}}
\institute{
     \inst{1} Institut f\"ur Theoretische Physik und Synergetik,
              Universit\"at Stuttgart\\ D-70550 Stuttgart, Germany\\
     \inst{2} Department of Chemistry, University of Southern California\\ 
              Los Angeles, CA 90089, USA}
\rec{}{}
\pacs{
\Pacs{05}{45$-$a}{Nonlinear dynamics and nonlinear dynamical systems}
\Pacs{03}{65Sq}{Semiclassical theories and applications}
      }
\maketitle
\begin{abstract}
Periodic orbit quantization requires an analytic continuation of 
non-convergent semiclassical trace formulae.
We propose a method for semiclassical quantization based upon
the Pad\'e approximant to the periodic orbit sums.
The Pad\'e approximant allows the re-summation of the typically
exponentially divergent periodic orbit terms.
The technique does not depend on the existence of a symbolic dynamics 
and can be applied to both bound and open systems.
Numerical results are presented for two different systems with chaotic
and regular classical dynamics, viz.\ the three-disk scattering system 
and the circle billiard.
\end{abstract}

Semiclassical theories link the spectrum of a quantum system to the
dynamics of its classical counterpart and thereby play an important
r\^ole in the deeper understanding of the relation between quantum
and classical mechanics.
Of particular interest are periodic orbit theories which describe the
quantum mechanical density of states in terms of contributions from the 
periodic orbits of the classical system.
The semiclassical trace formulae have been derived by Gutzwiller for
classically chaotic systems \cite{Gut67,Gut90} and by Berry and Tabor
for regular (integrable) systems \cite{Ber76}.
A common feature of these formulae is that in most cases the 
periodic orbit sum does not converge in those regions where the 
semiclassical eigenenergies or resonances are located. These resonances
are given as the poles of the periodic orbit sum,
\begin{equation}
 g(E) = \sum_{\rm po}
   {\cal A}_{\rm po}(E)e^{i[S_{\rm po}(E)/\hbar-{\pi\over 2}\mu_{\rm po}]} \; ,
\label{g_def:eq}
\end{equation}
where ${\cal A}_{\rm po}$, $S_{\rm po}$, and $\mu_{\rm po}$ are the amplitude,
classical action and Maslov index of the periodic orbit (po), respectively.
Formulae for the amplitudes ${\cal A}_{\rm po}$ are given in 
refs.\ \cite{Gut90,Ber76}.

In the last decade various  techniques have been developed  to circumvent the 
convergence problem of periodic orbit theory.
Examples are the cycle expansion technique \cite{Cvi89,Eck93,Eck95,Wir99},
the Riemann-Siegel type formula and pseudo-orbit expansions \cite{Ber90},
and a periodic orbit quantization rule for the counting function of
eigenvalues \cite{Aur92}.
These techniques have been proven to be very efficient for systems with 
special properties, {\it e.g.\/}, the cycle expansion for hyperbolic systems 
with an existing symbolic dynamics, and the other methods for the calculation 
of bound spectra.
Recently, a method for periodic orbit quantization has been
introduced \cite{Mai97,Mai98,Mai99a} which is based on the high resolution
analysis (harmonic inversion) of the semiclassical recurrence signal.
The method requires the knowledge of the periodic orbits up to a given 
maximum period (classical action), which depends on the mean density of states
({\it i.e.\/}, the Heisenberg time), and can be applied to open or bound 
systems with regular, chaotic, or even mixed \cite{Mai99b} classical dynamics.

In this Letter we propose an alternative method for periodic orbit 
quantization. It is the Pad\'e approximant (PA) to slowly convergent 
and/or divergent periodic orbit sums. In the former
or the latter case, the PA either significantly increases the 
convergence rate or analytically continues the {\it exponentially}
divergent series. The PA is especially robust for re-summing 
diverging series in many applications in
mathematics and theoretical physics \cite{Bak75}.
An important example is the summation of the divergent Rayleigh-Schr\"odinger
quantum mechanical perturbation series for, {\it e.g.\/}, atoms in 
electric \cite{CiV82} and magnetic \cite{Bel89} fields.
In periodic orbit theory the PA has been applied to
cycle-expanded Euler products and dynamical zeta functions \cite{Eck93}.
However, to the best of our knowledge, the PA has not yet been used for 
the direct summation of periodic orbit terms.

\section{Pad\'e approximant (PA) to the periodic orbit sums}
The PA to a complex function $f(z)$ is defined as a ratio of two polynomials 
and can be computed from the coefficients $a_n$ of the Maclaurin expansion 
of $f(z)$, {\it i.e.\/}, a power series $f(z)=\sum_{n=0}^\infty a_nz^n$
with finite or even zero radius of convergence in $z$ \cite{Bel89}.
However, eq.\ (\ref{g_def:eq}) does not have the functional form of a
Maclaurin power series expansion of $g(E)$ in the energy $E$.
Even disregarding this limitation, a direct computation of the PA to the 
sum in eq.\ (\ref{g_def:eq}) would be numerically unstable due to the 
typically large number of periodic orbit terms. Nevertheless,
considering $E$ as a parameter, $g(E)$ can be rearranged and written as 
a formal power series in an auxiliary variable $z$,
\begin{equation}
 g(z;E) = \sum_n (-iz)^n \left[ \sum_{\mu_{\rm po}=n}
   {\cal A}_{\rm po}(E)e^{iS_{\rm po}(E)/\hbar} \right] 
\; \equiv \; \sum_n a_n(E)z^n \; ,
\label{g_z_def:eq}
\end{equation}
where the maximal value of $n$ required for convergence of the PA is 
relatively small compared with the number of periodic orbit terms.
Of course, the arrangement (\ref{g_z_def:eq}) of the periodic orbit sum as a
power series is not unique, and expansions similar to eq.~(\ref{g_z_def:eq})
can be used for other ordering parameters $n$ of the orbits, {\it e.g.\/}, 
the cycle length in systems with an existing symbolic code.
However, if no symbolic dynamics exists, the sorting of orbits by their 
Maslov index is natural both physically and as a way to introduce an integer 
summation index and will be justified below by the successful numerical 
application of the method to a regular system without symbolic dynamics.
(For a discussion of the ordering of orbits in cycle-expansion techniques
see \cite{Chr92}.)
Note that the PA to the periodic orbit sum cannot be applied
without any ordering parameter, {\it i.e.\/}, when no symbolic dynamics 
exists and all Maslov indices are zero, which is the case, {\it e.g.\/},
for the Riemann zeta function as a mathematical model for periodic orbit 
quantization \cite{Mai98}.
The true value $g(E)$ of the periodic orbit sum is obtained by setting 
$z=1$ in eq.~(\ref{g_z_def:eq}), {\it i.e.\/}, $g(E)=g(1;E)$. 
In such a case, we have a point PA which is given as a ratio of two 
polynomials in $z$ whose coefficients are non-polynomial functions of $E$ 
all at a fixed value of $z$.
The usual implementation of the PA as a ratio of two polynomials in $z$
whose coefficients are computed, {\it e.g.\/}, via the Longman algorithm 
\cite{Lon79} would be advantageous if $g(z;E)$ were required for many values 
of $z$.
In the present case, at each given energy $E$ only one fixed value $z=1$ is 
needed and the PA is most efficiently computed by means of the recursive 
Wynn $\varepsilon$-algorithm \cite{Wyn56}. 

To briefly describe the $\varepsilon$-algorithm, we introduce a sequence of 
partial sums $\{A_n\}$ which converges to (or diverges from) its limit $A$ 
as $n\to\infty$. In the case of divergence, $A$ is called the `anti-limit' of
$\{A_n\}\,,$ as $n\to\infty$. Further, let $F$ be a transformation which maps 
$\{A_n\}$ into another sequence $\{B_n\}$.
The mapping $F$ will represent an accelerator, {\it i.e.\/}, sequence
$\{B_n\}$ will converge to the same limit $A$ faster than $\{A_n\}$ if the 
following condition is fulfilled: $(B_n-B)/(A_n-A)\to 0$, as $n\to\infty$.
In addition, the same $F$ can be applied to wildly divergent sequences 
$\{A_n\}$. Only non-linear mappings can simultaneously accomplish both goals 
to accelerate slowly convergent and induce convergence into divergent 
sequences. The transformation $F$ will be non-linear if its coefficients 
depend on $A_n$, {\it e.g.\/}, the so-called $e$-algorithm of Shanks 
\cite{Sha55}, whose mapping $F$ is the operator $e_k$ which converts sequence 
$\{A_n\}$ into $\{B_n\}$ via $e_k(A_n)=B_{k,n}=[n+k/k]$ ($n\ge 0$, $n\ge k$). 
This is the well-known Aitken $\Delta^2$-iteration process 
({\it i.e.\/}, the simplest PA, [1/1]) extended to higher orders $k$. 
The general term in the $k$th-order transform $B_{k,n}$ of $A_n$ can be 
computed efficiently from the stable and recursive $\varepsilon$-algorithm of
Wynn \cite{Wyn56}, {\it i.e.\/}, $e_s(A_m)=\varepsilon_{2s}^{(m)}=[m+s/s]$, 
where
\begin{equation}
   \varepsilon_{s+1}^{(m)}
 = \varepsilon_{s-1}^{(m+1)}
 + 1/\left(\varepsilon_s^{(m+1)}-\varepsilon_s^{(m)}\right) \; ; \quad
   m, s \ge 0
\label{eps_alg:eq}
\end{equation}
with $\varepsilon_{-1}^{(m)}=0$, $\varepsilon_0^{(m)}=A_m$,
$\varepsilon_{2s+1}^{(m)}=1/e_s(\Delta A_m)$ and where $\Delta$ is the forward
difference operator: $\Delta x_j=x_{j+1}-x_j$.
When $\{A_n\}$ is the sequence of partial sums of a power series, the
two-dimensional array $\varepsilon_{2s}^{(m-s)}$ yields the upper half of the
well known Pad\'e table $[m/s]$. However, the $\varepsilon$-algorithm
need not necessarily be limited to power series.

The procedure to apply the above PA to semiclassical quantization 
by summation of periodic orbit terms is as follows.
For a given system we calculate the periodic orbits up to a chosen maximum
ordering parameter $n \le n_{\rm max}$ where $n$ can be but is not
necessarily related to the Maslov indices of orbits.
Note that this set of orbits usually differs from the set of orbits
with classical action $S_{\rm po}\le S_{\rm max}$, which is required for
periodic orbit quantization by harmonic inversion \cite{Mai97,Mai98,Mai99a}.
From the quantities ${\cal A}_{\rm po}$, $S_{\rm po}$ and $\mu_{\rm po}$ of
these orbits, we compute the partial sums 
({\it e.g.\/}, with $n\le \mu_{\rm max}$ the Maslov index):
\begin{equation}
 A_n = \sum_{\mu_{\rm po}\le n} {\cal A}_{\rm po}(E)
    e^{i[S_{\rm po}(E)/\hbar-{\pi\over 2}\mu_{\rm po}]} \; .
\label{A_n:eq}
\end{equation}
The sequence $\{A_n\}$ of partial sums is used as input to the
$\varepsilon$-algorithm, eq.\ (\ref{eps_alg:eq}), to obtain
a converged value $g(E)$ of the periodic orbit sum (\ref{g_def:eq}).
The semiclassical eigenenergies or resonances are given as the poles of $g(E)$ 
and are obtained by searching numerically  for the zeroes
of the reciprocal function $1/g(E)$. 
Such a search requires the evaluation of, {\it e.g.\/}, 
${\cal A}_{\rm {po}}(E)$ and $S_{\rm {po}}(E)$ 
at complex values of $E$. This is straightforward for the scaling systems
considered in this Letter. The number of the poles of $g(E)$ is not
constrained by the size of the sequence $\{A_n\}$ of partial sums since
our PA is a ratio of two non-polynomial functions of $E$.
We now demonstrate the power of the method on two physical
examples with completely different dynamical properties, viz.\
the open three-disk billiard and the bound circle billiard.

\section{The three-disk scattering system}
As the first example we consider a billiard system consisting of three
identical hard disks with unit radii, $R=1$, displaced from each other
by the same distance $d$.
This simple, albeit non-trivial, scattering system has served as a
model for the development of the cycle expansion method 
\cite{Cvi89,Eck93,Eck95,Wir99} and periodic orbit quantization by
harmonic inversion \cite{Mai97,Mai98,Mai99a}.
The three-disk scattering system is invariant under the symmetry operations
of the group $C_{3v}$, {\it i.e.\/}, three reflections at symmetry lines 
and two rotations by $2\pi/3$ and $4\pi/3$.
Resonances belong to one of the three irreducible subspaces $A_1$, $A_2$, 
and $E$ \cite{Cvi93}.
In the following we concentrate on the resonances of the subspace $A_1$.
For $d=6$, semiclassical resonances were calculated by 
the cycle expansion technique, including periodic orbits up to cycle length
$n=13$ \cite{Eck95,Wir99}.
In order to demonstrate the power of the PA
we first apply it to the previously studied case with $R=1$, $d=6$.
In billiards, which are scaling systems, the shape of periodic orbits
does not depend on the energy $E$, 
and the classical action is given by 
the length $L$ of the orbit ($S_{\rm po}=\hbar kL_{\rm po}$), where 
$k=|{\bf k}|=\sqrt{2ME}/\hbar$ is the absolute value of the wave vector 
to be quantized.

\begin{figure}
\vspace{5.0cm}
\includegraphics{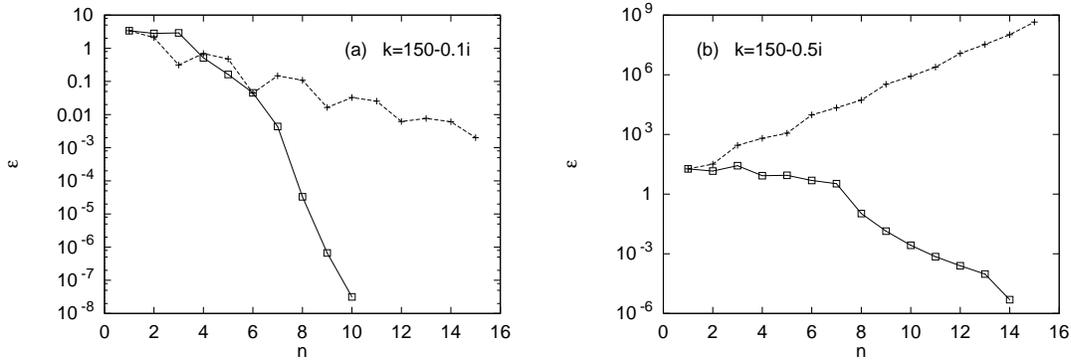}
\caption{
Convergence vs.\ exponential divergence of the partial periodic orbit sums 
for the three-disk scattering system with $R=1$, $d=6$ as functions of 
the order $n$ at (a) complex wave number $k=150-0.1i$ and (b) $k=150-0.5i$.
Dashed lines and plus symbols: 
Error values $\varepsilon=|A_n-A_{15}^{\rm PA}|$ for the sequence $A_n$ 
without Pad\'e approximation.
Solid lines and squares: 
Error values $\varepsilon=|A_n^{\rm PA}-A_{15}^{\rm PA}|$ for the Pad\'e 
approximant $A_n^{\rm PA}$ to the periodic orbit sums.
}
\label{fig1}
\end{figure}

We have calculated all periodic orbits with Maslov index $\mu_{\rm po}\le 30$,
which corresponds to the set of orbits with cycle length $n\le 15$.
The resulting sequence of the partial sums $\{A_n\}$ of periodic orbit 
terms (eq.\ \ref{A_n:eq}) converges for wave numbers $k$ above the
borderline ${\rm Im}~k = -0.121\,557$ \cite{Cvi89} which separates the 
domain of absolute convergence of the periodic orbit sum from the domain 
where analytic continuation is necessary, but strongly diverges deep in 
the complex plane, where the resonance poles are located.
This is illustrated in fig.~\ref{fig1} for two different wave numbers $k$.
The dashed line and the plus symbols in fig.~\ref{fig1}a show the
convergence of the sequence $\{A_n\}$ at $k=150-0.1i$.
What is plotted is the error values $\varepsilon=|A_n-A_{15}^{\rm PA}|$,
with $A_{15}^{\rm PA}$ the best known approximation to the true limit
of the periodic orbit sum.
As can be seen this sequence is slowly convergent, and about three
significant digits are obtained at $n=15$.
The convergence can be accelerated using the PA, as is seen by the 
solid line and squares in fig.~\ref{fig1}a showing the error values 
$\varepsilon=|A_n^{\rm PA}-A_{15}^{\rm PA}|$ for the sequence of the 
Pad\'e approximants $\{A_n^{\rm PA}\}$ to the periodic orbit sum.
The periodic orbit sum has converged to six significant digits already by
$n=9$.
The situation is much more dramatic in the deep complex plane
(${\rm Im}~k<-0.122$), {\it e.g.\/}, at $k=150-0.5i$.
Here, the sequence of the partial sums $\{A_n\}$ of periodic orbit terms 
exhibits {\em exponential} divergence, as can be seen by the dashed line
and plus symbols in fig.~\ref{fig1}b.
Nevertheless, this sequence converges when subjected to the PA implemented
through the $\varepsilon$-algorithm (see the solid line and squares in 
fig.~\ref{fig1}b).

\begin{figure}
\vspace{4.0cm}
\includegraphics{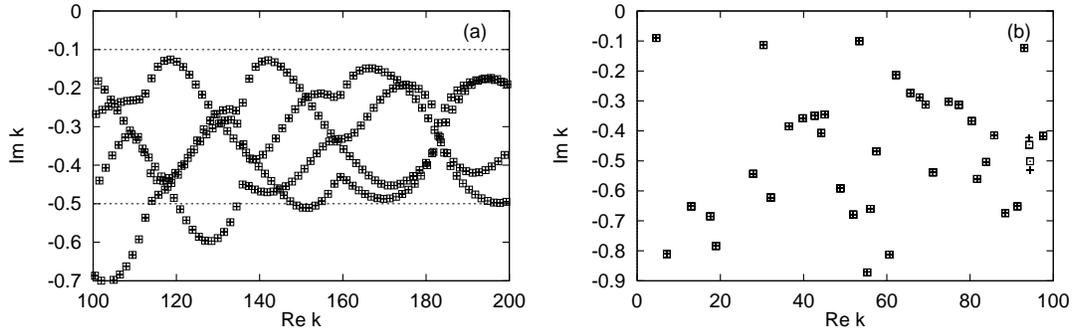}
\caption[]{
(a) Semiclassical resonances ($A_1$ subspace) for the three-disk scattering  
system with $R=1$, $d=6$. 
The crosses and squares mark the semiclassical resonances obtained 
by Pad\'e approximant to the periodic orbit sum and by cycle expansion 
\protect{\cite{Wirz}}, respectively. 
The convergence properties of the periodic orbit sum and its Pad\'e 
approximant are discussed at ${\rm Im}~k=-0.1$ and ${\rm Im}~k=-0.5$ 
marked by the dashed lines (see text). 
(b) Same as (a) but with small distance $d=2.5$ between the three disks. 
}
\label{fig2}
\end{figure}

The resonances of the three-disk scattering systems have been obtained
by a numerical two-dimensional root search in the complex $k$-plane for 
the zeroes of the function $1/g(k)$, where $g(k)$ is the PA to the periodic
orbit sum. A typical subset of the semiclassical resonances obtained is 
presented as crosses in fig.~\ref{fig2}a.
They agree perfectly with results obtained by the cycle expansion method
in refs.\ \cite{Eck95,Wir99}.
For comparison the results of the cycle expansion \cite{Wirz} are presented
as squares.
The dashed lines in fig.~\ref{fig2}a denote the imaginary parts 
${\rm Im}~k=-0.1$ and ${\rm Im}~k=-0.5,$ for which the convergence properties
of the periodic orbit sum and the PA has been discussed above.

Calculations similar to those discussed above for $d=6$ have also been 
carried out at a shorter distance, $d=2.5$, between the three disks.
The semiclassical resonances obtained by PA to the periodic orbit sum 
and by cycle expansion \cite{Wirz} are presented as crosses and squares 
respectively in fig.~\ref{fig2}b.
Again, the agreement is excellent with the exception of minor discrepancies
for the imaginary parts of two resonances with ${\rm Re}~k\approx 94.3$.
It should be noted that the rate of convergence of the PA decreases
when the distance $d$ between the disks is reduced as is also the case
when cycle-expansion methods are applied.
A comparison of the convergence properties of the PA and the cycle-expansion
for this specific system will be published elsewhere \cite{Eckhardt_comment}.

\section{The circle billiard}
\begin{figure}
\vspace{4.0cm}
\includegraphics{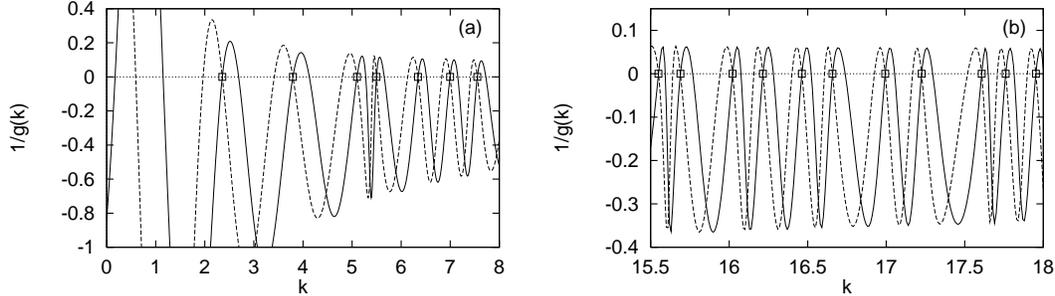}
\caption{
Real part (solid line) and imaginary part (dashed line) of the function
$1/g(k)$ for the circle billiard with radius $R=1$ obtained by Pad\'e 
approximant to the periodic orbit sum.
The zeroes agree perfectly with the exact positions of the semiclassical
eigenvalues (from eq.~\ref{EBK}) marked by the squares.
}
\label{fig3}
\end{figure}

We now demonstrate the PA to periodic orbit sums for the example of the 
circle billiard.
In sharp contrast to the three-disk system analyzed above, the circle billiard
is an integrable and bound system, and, to the best of our knowledge, has not
yet been treated by the cycle expansion technique \cite{Cvi89} or pseudo-orbit 
expansion \cite{Ber90}.
The exact quantum mechanical eigenvalues $E=\hbar^2k^2/2M$ of the circle
billiard are given by zeroes of Bessel functions $J_{|m|}(kR)=0$, where 
$m =0,\pm 1,\pm2,,\dots$ 
is the angular momentum quantum number and $R$ is the radius of the circle.
The semiclassical eigenvalues can be obtained by an Einstein-Brillouin-Keller
(EBK) torus quantization \cite{Per77} resulting in the quantization 
condition
\begin{equation}
 kR\sqrt{1-(m/kR)^2} - |m|\arccos{|m|\over kR} = \pi\left(n+{3\over 4}\right)
\label{EBK}
\end{equation}
where $n=0,1,2,\dots$ is the radial quantum number.
In the following we choose $R=1$.
The periodic orbits of the circle billiard are those orbits for which the 
angle between two bounces is a rational multiple of $2\pi$, {\it i.e.\/}, 
the periods $L_{\rm po}$ are obtained from the condition
\begin{equation}
 L_{\rm po} = 2m_r \sin \gamma \; ,
\end{equation}
with $\gamma\equiv\pi m_\phi/m_r$. Here, $m_\phi=1,2,\dots$ is the number 
of turns of the orbit around the centre of the circle and 
$m_r=2m_\phi,2m_\phi+1,\dots$ is the number of reflections at the boundary 
of the circle.
Periodic orbits with $m_r\ne 2m_\phi$ can be traversed in two directions
and thus have multiplicity 2.
For the amplitudes ${\cal A}_{\rm po}$ of the circle billiard, the
Berry-Tabor formula \cite{Ber76} for integrable systems yields
\begin{equation}
   {\cal A}_{\rm po}
 = \sqrt{-i\pi/2}L_{\rm po}^{3/2}/m_r^2 \; ,
\label{A_circ:eq}
\end{equation}
and the Maslov index is
\begin{equation}
 \mu_{\rm po} = 3m_r  \; .
\label{Maslov_circ:eq}
\end{equation}
It follows from here that the partial periodic orbit sums $A_n$ in 
eq.~(\ref{A_n:eq}) should be calculated with order $n=m_r$.
(The choice $n=\mu_{\rm po}=3m_r$ is also possible but the numerical
calculations would be less robust because the order of the polynomials
in the PA are increased by a factor of 3.)
We included all periodic orbits with $m_r<100$ in the PA to the function 
$g(k)$.
The real and imaginary parts of $1/g(k)$ are presented as solid and
dashed lines respectively in fig.~\ref{fig3}.
The zeroes of the function $1/g(k)$ agree perfectly to at least
seven significant digits with the exact positions of the
semiclassical eigenvalues obtained from eq.~(\ref{EBK}) and marked
by the squares in fig.~\ref{fig3}.

For chaotic systems the efficiency of semiclassical periodic orbit
quantization can be estimated by comparing the length of the longest
orbit included in the calculation with the Heisenberg period
$L_H=2\pi\bar\rho$ given by the average density of states.
However, for integrable systems the nearest neighbour spacing statistics
is a Poisson distribution with high probability of nearly degenerate
levels.
The lengths of periodic orbits should therefore be compared with the 
length $L_{\rm max}=2\pi/\Delta k_{\rm min}$ (with $\Delta k_{\rm min}$ 
the smallest level spacing) rather than with the Heisenberg length.
For the circle billiard, using orbits with period $L_{\rm po}<200$, it 
is no problem to resolve level spacings $\Delta k<0.01$, {\it i.e.\/}, 
spacings where the characteristic length 
$L_{\rm max}=2\pi/\Delta k \approx 628$ is by more than a factor of three 
larger than the maximum length of the periodic orbits.\\

In conclusion, we have introduced the Pad\'e approximant to perform 
summation of periodic orbit terms as a simple but powerful method for 
periodic orbit quantization.
The Pad\'e approximant allows the re-summation of the typically 
exponentially divergent terms of the semiclassical trace formulae.
The method has been demonstrated on two systems with completely different 
classical dynamics, viz.\ the classically chaotic three-disk scattering 
problem and the integrable circle billiard.
The Pad\'e approximant can be applied when the total periodic orbit sum can be
divided into partial sums with respect to an integer ordering parameter $n$.
It is an interesting speculation if the latter might be related to the 
cycle length of periodic orbits if a symbolic dynamics exists, or, in the 
general case, to the Maslov indices of orbits.
Clearly, an in-depth investigation of this point will be worthwhile but
goes beyond the scope of this Letter.

\stars
We thank G.~Wunner and B.~Eckhardt for fruitful discussions and
comments on the manuscript.
Financial support from the National Science Foundation 
(NSF-Grant No.\ PHY-9802534) is gratefully acknowledged.
JM thanks HST for the kind hospitality during his visit at the University
of Southern California where this work was initiated.


\end{document}